\documentclass{article}
\usepackage{spconf,amsmath,graphicx}
\usepackage{booktabs}
\usepackage{multirow}
\usepackage{bigstrut}

\title{THE USTC-Ximalaya SYSTEM FOR THE ICASSP 2022 MULTI-CHANNEL MULTI-PARTY MEETING TRANSCRIPTION (M2MeT) CHALLENGE}
\name{\begin{tabular}{c}Maokui He$^1$, Xiang Lv$^2$, Weilin Zhou$^2$, JingJing Yin$^2$, Xiaoqi Zhang$^1$, Yuxuan Wang$^1$\\Shutong Niu$^1$, Yuhang Cao$^2$, Heng Lu$^2$, Jun Du$^1$\sthanks{This work was supported by the National Natural Science Foundation of China under Grants No. 62171427}, Chin-Hui Lee$^3$\end{tabular}}

\address{$^1$University of Science and Technology of China, HeFei, China\\$^2$Ximalaya Inc., ShangHai, China\\$^3$Georgia Institute of Technology, Atlanta, GA, USA}

\begin{document}
%
\maketitle

\begin{abstract}
 We propose two improvements to target-speaker voice activity detection (TS-VAD), the core component in our proposed speaker diarization system that was submitted to the 2022 Multi-Channel Multi-Party Meeting Transcription (M2MeT) challenge.  These techniques are designed to handle multi-speaker conversations in real-world meeting scenarios with high speaker-overlap ratios and under heavy reverberant and noisy condition. First, for data preparation and augmentation in training TS-VAD models, speech data containing both real meetings and simulated indoor conversations are used. Second, in refining results obtained after TS-VAD based decoding, we perform a series of post-processing steps to improve the VAD results needed to reduce diarization error rates (DERs). Tested on the ALIMEETING corpus, the newly released Mandarin meeting dataset used in M2MeT, we demonstrate that our proposed system can decrease the DER by up to 66.55/60.59\% relatively when compared with classical clustering based diarization on the Eval/Test set.
\end{abstract}


\section{Introduction}
\label{sec:intro}
Speaker diarization is the task to address the "who speaks at when" problem in multi-speaker conversations. It plays an important role in many applications, primarily of automatic speech recognition (ASR), e.g., for meeting transcription. Besides, it can also provide priors for speaker activity mask based separation, e.g., guided source separation \cite{kanda2019guided} for CHiME6 diner party ASR. Conventional clustering-based methods, which mainly include speech activity detection (SAD), speech segmentation, speaker feature extraction, and speaker clustering, are widely used in speaker diarization \cite{CD2}. However, this framework inherently makes an assumption that every segment can only be assigned with a single speaker label through hard clustering. Re-segmentation \cite{sell2015diarization} is often adopted to handle speech-overlap segments, but overlap detection is still challenging task in most situations.

In order to directly solve the overlap problem and reduce the diarization errors, such as end-to-end neural speaker diarization (EEND) \cite{fujita2019end, fujita2019end2} and target-speaker voice activity detection (TS-VAD) \cite{tsvad_Medennikov} were proposed. They judged each speaker’s activeness for each frame, so they can fundamentally estimate multiple speakers at the same time. But the limitation lies in the total number of speakers is fixed. Further research was taken on handling flexible number of speakers on EEND \cite{horiguchi2020end}. And \cite{he21c_interspeech} provides a strategy for handling an unknown numbers of multiple speakers of TS-VAD. Based on the second network-based method, we explore the effect of training data augmentation and post-processing on TS-VAD method with an unknown number of multiple speakers.

In M2MeT constrained training data condition, only the ALIMEETING corpus \cite{yu2021m2met}, AISHELL-4 \cite{AISHELL42021} and CN-CELEB \cite{fan2020cn} are allowed to train the diarization model. We did a series of experiments on the different training data setups including ALIMEETING original far-field data, dereverbed ALIMEETING data, AISHELL-4 original and dereverbed data, simulated meeting-like data with CN-CELEB. Experiments show that the diarization performance is improved with increasing the amount of training data. Then, we performed post-processing on the best TS-VAD model we trained. Merging the two segments with short silence interval for each speaker addresses the problem that TS-VAD may predict the speaker brief pauses frames as nonspeech. To fuse the above results with reliable golden speech segments, we deleted the silent speech segments in oracle VAD to reduce false alarm speech and labeled the silent segments that are speech in oracle VAD with the longest talking person nearby to reduce missed detection speech. DOVER-Lap \cite{Raj2021ReformulatingDL} of multi-channel TS-VAD results could also bring some improvement. Finally, re-decoding with the i-vectors extracted from TS-VAD results also brings a little improvement. Our best result achieved DER 7.80/9.14\% on the Eval/Test set.


\section{System Description}
\label{sec:System_Description}
\begin{figure}[t]
 \begin{center}
 \includegraphics[width=1.0\linewidth]{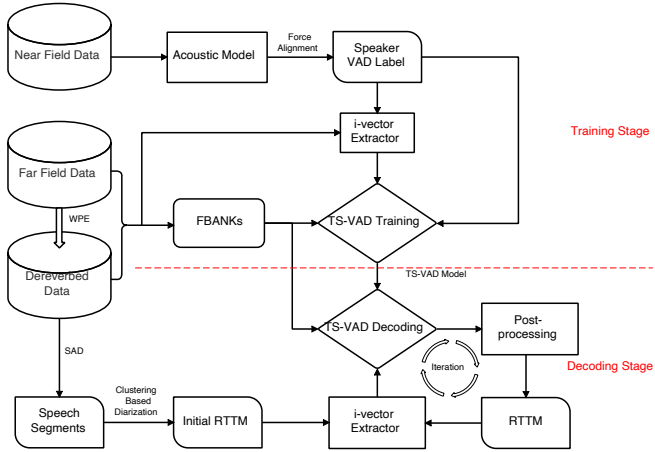}
 \caption{Our speaker diarization system for M2MeT Challenge.}
\label{fig:overall_structure}
 \end{center}
\end{figure}

Figure~\ref{fig:overall_structure} illustrates our overall speaker diarization system for the 2022 M2MeT challenge. The core technology is that we used TS-VAD with an unknown number of multiple speakers \cite{he21c_interspeech} and tried some new strategies for the multi-channel Mandarin meeting scenario with heavy reverb and noise. In the training stage, the training data for TS-VAD will be introduced in Section~\ref{sec:data_preparation}. In decoding stage, we decoded all far-field channel data with the estimated i-vectors from the clustering-based speaker diarization (CSD) and performed a series of post-processing strategies including thresholding, merging two segments with short silence interval, fusing results with golden speech segments (oracle VAD is provided) and DOVER-Lap \cite{Raj2021ReformulatingDL} of the multi-channel systems. Moreover, we also estimate i-vectors with the TS-VAD results which are more reliable than CSD and repeat the whole decoding process.

\subsection{Clustering-Based Speaker Diarization}
Before decoding with TS-VAD, we need an initial diarization result to get each speaker's segments for extracting corresponding i-vectors. M2MeT baseline \cite{yu2021m2met} provides AHC with Variational Bayesian HMM clustering (VBx) \cite{landini2022bayesian}. First, for the speaker embedding network, we replace the baseline ResNet with ECAPA-tdnn(C=512) \cite{desplanques2020ecapa}. We split the training data into 1.5s segments, and extracted 80 dimension log mel filter-banks as the model input. We train 5 epochs with no data augmentation for a fast convergence. Then we train another 5 epochs with 6 fold data augmentation using noise/reverb from openrir \cite{park2019auto} and MUSAN \cite{snyder2015musan} to improving embedding performances. We train the ECAPA embedding network using cosine similarity directly, which save us from the effort of training an extra PLDA model. Second, for clustering, we use spectral clustering with cosine similarity, and use an automatic selection of binarization threshold to determine the number of speakers \cite{park2019auto}. 

\subsection{TS-VAD with an Unknown Number of Speakers}
The performance of CSD is good enough to deal with most cases, but it can’t well handle overlapping speech which leads to a high miss in diarization error rate (DER). Here we adopt TS-VAD to cope with over-lapped regions in each recording and further reduce the chance of misjudging speakers (speaker error in DER). The original TS-VAD model \cite{tsvad_Medennikov} takes conventional speech features (e.g. log Mel filter-banks (FBANKs)) as input, along with i-vectors corresponding to each speaker, and predicts per-frame speech activities for a fixed number of speakers simultaneously, which directly handles overlapping problems. In the flexible number of speakers case \cite{he21c_interspeech},  the number of output nodes $N$ is chosen as the maximum number of speakers in any recording in the training set, which is 4 for the ALIMEETING whose speaker number of each recording ranges from 2 to 4. First, the number of speakers $\hat{N}$ in each recording is estimated according to the oracle label when training and  a CSD system when decoding. If this estimate $\hat{N}$ is equal to the number of output nodes $N$, then no further effort is required and the trained model can be directly applied to the recording. If $\hat{N}$ is larger than $N$, select $N$ out of the estimated speakers who have the longest non-overlapping speaking duration in the initial diarization output (oracle label for training and CSD results for decoding), and discard others. If $\hat{N}$ is smaller than $N$, assign $\hat{N}$ of the $N$ output nodes to these “test” speakers, and assign the remaining $N - \hat{N}$ nodes to dummy speakers selected from the training set randomly. These dummy training speakers are abandoned when generating the final diarization output. The input number of speakers can be fixed to $N$ with this strategy for both training and decoding. The input speaker i-vectors are extracted with oracle speaker segments in the training stage and CSD results in the decoding stage.

\subsection{TS-VAD Decoding and Post-processing}
With the CSD result, we extracted the i-vectors of speaker in each session and combined them with FBANKs computed with original data for TS-VAD models were trained only with ALIMEETING\_Train\_RAW and dereverbed data for the other TS-VAD models as the model input. Then we got each speaker's frame-level speech/nonspeech probabilities. In the post-processing stage, we first performed thresholding with the probabilities and produced a preliminary result. Then, we merged the two segments with short silence interval for each speaker to address the problem that TS-VAD may predict the speaker brief pauses frames as nonspeech, which are labeled as speech in the manual label. Then, we fused the results with golden speech segments (oracle VAD) by deleting the speech segments that are silence in oracle VAD and labeling the silent segments with the longest talking person nearby that are speech in oracle VAD. Next, we did DOVER-Lap on the multi-channel results (8 systems). Finally, we estimated i-vectors with the TS-VAD results which are more reliable than CSD, and redid the above process and made 3 iterations. 

\section{TS-VAD Training Data Preparation}
\label{sec:data_preparation}
First,  without any data augmentation, 100-dimensions i-vectors was trained on CN-CELEB with 512 Gaussians in UBM. Our previous experiments showed data augmentation with adding noise and reverb led to worse results. Second, we performed a series of experiments to demonstrate the effect of each part of the data including ALIMEETING far-field multi-channel data, dereverbed ALIMEETING far-field  multi-channel data, simulated meeting-like data with CN-CELEB \cite{fan2020cn}  and AISHELL-4 \cite{AISHELL42021}.

\noindent\textbf{ALIMEETING far-field multi-channel data} is real meeting data recorded by 8-channel directional microphone array. It consists of 104.75 hours for training (Train), 4 hours for evaluation (Eval) and 10 hours as test set (Test). Each session consists of a 15 to 30-minute discussion by a group of participants. The meeting room sizes range from 8 to 55 $m^2$ and the microphone-speaker distance ranges from 0.3 to 5.0 m. The average speech overlap ratio of Train and Eval set are 42.27 \% and 34.76 \%, respectively. All those room settings are crucial parameters in the following simulating data part. Without directly using the manual label from the transcription  for training, we toke the force alignment as the final training label which deleted the silence of manual label segments caused by brief pauses of speakers and ensured a more accurate frame-level training targets. First, we followed the AISHELL-2 Mandarin ASR pipeline \cite{aishell2} and trained the GMM model with the ALIMEETING near-field data. Then, we got each speaker VAD label in each session with tri3 (LDA+MLLT) alignment on the corresponding speaker's headset microphone.

\noindent\textbf{Dereverbed ALIMEETING} is a dereverberation version of far-field multi-channel data. A well-known algorithm for removing the acoustic reflections is WPE \cite{yoshioka2012generalization}, which has been proven to improve the results significantly in various speech processing tasks \cite{delcroix2014linear, caroselli2017adaptive}. In WPE, the reverberant component in speech is firstly estimated based on the multi-channel linear prediction, and then removed from the observations in the maximum likelihood sense. In this meeting challenge, we used the open-sourced software NARA-WPE \cite{drude2018nara} to perform the dereverberation for each far-field channel data. The offline mode is used due to its superior performance upon the online setting. Same as before, the force alignment label is taken as the training label of dereverbed data.

\noindent\textbf{Simulated meeting-like data with CN-CELEB} is a room simulated conversation data. We got a conversation session by randomly selecting 2 to 4 speakers from CN-CELEB and their 1 to 10 sentences and combining those utterances with an overlap ratio from 0 to 40\%. Then, we performed the synthetic room impulse response (RIR) for this session using the image method \cite{allen1979image}, where the room size (length, width, height) ranges from (2m, 2m, 2.5m) to (10m, 10m, 4.5m) while the distance between speaker and microphone ranges from (0.3m, 0.3m, 0.01m) to (5m, 5m, 0,8m) and the RT60 ranges from 0.15 to 0.3. We also added the noise and reverb from openrir \cite{park2019auto} and MUSAN \cite{snyder2015musan} for data augmentation. To get a reliable training label, we performed WebRTC Voice Activity Detector (VAD) ~\footnote{https://github.com/wiseman/py-webrtcvad} on the original CN-CELEB utterance to remove the possible silence frames.

\noindent\textbf{AISHELL-4} is a sizable Mandarin speech corpus collected by an 8-channel circular microphone array for speech processing in conference scenarios. The dataset consists of 211 recorded meeting sessions, each containing 4 to 8 speakers, with a total of 120 hours. The same dereverbed procedure as before was also performed on AISHELL-4. Unlike ALIMEETING that provides the near-field data, we do nott get near-field data to perform forced alignment for AISHELL-4. We tried doing forced alignment on AISHELL-4 far-field data, but failed to get a reliable label set because some segments were recorded not clearly by the microphone array. So we directly used the given label set for training.

For description convenience, we adopt the following abrrevatios for the above four sets: ALIMEETING\_Train\_RAW, ALIMEETING\_Train\_WPE, CN-CELEB\_Simu, AISHELL-4 (including original  (RAW) and dereverbed (WPE) data). Table~\ref{tab:training_details} shows the details of all those training data.

\begin{table}
\small
\caption{TS-VAD train data details.}
\vspace{10pt}
\label{tab:training_details}
\centering
\begin{tabular}{lcc}
\toprule
\multicolumn{1}{c}{\textbf{Data}} & \multicolumn{1}{c}{\textbf{Duration (h)}} & \multicolumn{1}{c}{\textbf{Training label}} \\
\midrule
ALIMEETING\_Train\_RAW & 838 & Force alignment \\
ALIMEETING\_Train\_WPE  & 838 & Force alignment \\
CN-CELEB\_Simu  & 4978  & WebRTC VAD \\
 AISHELL-4  & 1720 & Manual \\
\bottomrule
\end{tabular}
\vspace{-10pt}
\end{table}

\section{Experiments and Result Analysis}
\label{sec:Experiments_Result}
\subsection{Experimental Setup}
The total number of parameters for 4-speaker TS-VAD is 35.6 M. During the training stage, each session was cut into short segments with 8s window length and 6s window shift. we performed mixup \cite{zhang2017mixup} within each session to ensure that the input i-vectors were not be changed. Adam optimizer \cite{kingma2017adam} was used to update model parameters and the learning rate was set to 0.0001. We trained TS-VAD models with pytorch on 4 GeForce RTX 3090s and batchsize was set to 128. It toke 4.63s to train one-hour data. During the decoding stage, we extracted one i-vector for each speaker in the whole session with his/her nonoverlap segments. Then, we fed the 8s short segments with the extracted speaker i-vectors into the TS-VAD model and got the speakers' speech/nonspeech probabilities. The accuracy of speaker diarization system in this track is measured by Diarization Error Rate (DER) \ where DER \cite{fiscus2006rich} is calculated as: the summed time of three different errors of speaker confusion (SC), false alarm (FA) and missed detection (MD) divided by the total duration time.

\subsection{Effect of Training Data Augumentation for TS-VAD}
\label{sec:data_augmentation}
The diarization results comparison among the training data setups are shown in Table 2. We only performed binarization with the threshold to generate the diarization results in this table. CSD was implemented with spectral clustering on x-vectors extracted with ECAPA-tdnn embedding network. The primary TS-VAD model trained only with ALIMEETING\_Train\_RAW and decoded with channel1 of ALIMEETING Eval original data has shown 4.37\% absolutely (abs.) DER decrease compared with CSD. The dereverbed data for both training and decoding gives another improvement. By training TS-VAD with both ALIMEETING\_Train\_RAW and ALIMEETING\_Train\_WPE, there is 1.34\% abs. decrease compared with the original model. The DER can be reduced further by adding new CN-CELEB\_Simu and AISHELL-4 data. Finally, we reduced DER from 23.14\% to 16.50\% compared CSD with the best TS-VAD model only with thresholding.

\begin{table}
\small
\label{tab:2}
\centering
\caption{DER(\%) performance comparison on the ALIMEETING Eval set among different training data setups.}
\vspace{10pt}
\begin{tabular}{lcccc}
\toprule
\multicolumn{1}{c}{\textbf{Training data}} & \multicolumn{1}{c}{\textbf{FA}} & \multicolumn{1}{c}{\textbf{MD}} & \multicolumn{1}{c}{\textbf{SC}} & \multicolumn{1}{c}{\textbf{DER}} \\
\midrule
CSD & 0.00 & 20.94 & 2.20 & 23.14 \\
\midrule
ALIMEETING\_Train\_RAW & 5.70 & 10.37 & 2.70 & 18.77 \\
ALIMEETING\_Train\_WPE & 6.49 & 9.62 & 1.99 & 18.09 \\
ALIMEETING (RAW+WPE) & 5.43 & 10.05 & 1.95 & 17.43 \\
\quad+CN-CELEB\_Simu & 6.18 & 9.19 & 1.53 & 16.89 \\
\qquad+AISHELL-4 & 5.29 & 9.68 & 1.52 & 16.50 \\
\bottomrule
\end{tabular}
\vspace{-10pt}
\end{table}

\subsection{Effect of Post-processing}
\label{sec:post-processing}
In the best TS-VAD model trained with ALIMEETING, CN-CELEB\_Simu and AISHELL-4, and decoded with dereverbed ALIMEETING, we show the diarization performance on different post-processing strategies in Table 3. Firstly, merging the two segments with short silence interval for each speaker bring huge performance improvement compared with only thresholding. MD was decreased significantly by assigning those short silence segments in thresholding results. FA was also reduced because the best threshold for T+M is larger than T. Fusing the above results with golden speech segments can significantly decrease FA/MD with only a little SC increase. Then DOVER-Lap of the 8 channel results can bring another improvement. Finally, we re-estimated i-vectors with the fusion results and achieved best DER 7.80\% on the Eval set with the iteration strategy three times.
\begin{table}
\small
\label{tab:effect_post_processing}
\centering
\caption{DER(\%) performance comparison on the Eval set with different post-processing. T, M, and F stand for binarization with the threshold, merging two segments with short silence interval, and fusing TS-VAD results with golden speech segments, respectively.}
\vspace{10pt}
\begin{tabular}{lcccc}
\toprule
\multicolumn{1}{c}{\textbf{Post-processing}} & \multicolumn{1}{c}{\textbf{FA}} & \multicolumn{1}{c}{\textbf{MD}} & \multicolumn{1}{c}{\textbf{SC}} & \multicolumn{1}{c}{\textbf{DER}} \\
\midrule
T & 5.29 & 9.68 & 1.52 & 16.50 \\
T + M & 4.65 & 6.89 & 1.50 & 13.04 \\
T + M + F & 2.63 & 5.71 & 1.81 & 10.14 \\
T + M + F + DOVER-Lap & 2.72 & 4.51 & 1.23 & 8.46 \\
\quad+ Iteration & 2.79 & 3.92 & 1.09 & 7.80 \\
\bottomrule
\end{tabular}
\vspace{-10pt}
\end{table}

\subsection{Analysis of Final Results}
Table 4 shows our final results on both Eval and Test sets. We first made a slight improvement with CSD compared with the official baseline. We found a DER gap of about 1.2\% between the Eval and Test set. In detail, the system performed "poor" on the Test set where the high overlap ratio sessions with 4 speakers appear more frequently.
\begin{table}
\small
\label{tab:final_results}
\centering
\caption{The final DER(\%) performance comparison on the ALIMEETING Eval and Test results.}
\vspace{10pt}
\begin{tabular}{cccc}
\toprule
\multirow{2}{*}{\textbf{System}} & \textbf{\#Speaker} & \textbf{Eval} & \textbf{Test} \\
& \textbf{Collar size (s)} & \textbf{0.25 / 0} & \textbf{0.25 / 0} \\
\bottomrule
Baseline & 2,3,4 & 15.24 / - & 15.60 / - \\
\midrule
CSD & 2,3,4 & 14.03 / 23.14 & 14.22 / 23.19 \\
\midrule
\multirow{4}{*}{TS-VAD} & 2,3,4 & 2.82 / 7.80 & 4.05 / 9.14 \\
& 2 & 0.47 / 3.29 & 0.44 / 2.84 \\
& 3 & 3.43 / 8.13 & 3.67 / 7.18 \\
& 4 & 4.15 / 10.21 & 7.28 / 13.87 \\
\bottomrule
\end{tabular}
\vspace{-10pt}
\end{table}

\section{Conclusions}
\label{sec:Conclusions}
In this paper, we adopt TS-VAD with an unknown number of speakers for speaker diarization on ALIMEETING data. First, we found that combining ALIMEETING, CN-CELEB, AISHELL-4 and their dereverbed data in training gives the best TS-VAD results. Second, after each iteration of TS-VAD based decoding, a series of post-processing strategies can be utilized to further refine detection of speaker and silence segments, leading to big diarization performance improvements. In the future, the effect of denoised data on diarization systems will also be explored.


\pagebreak
\small
\bibliographystyle{IEEEbib}
\bibliography{strings,refs}

\begin{thebibliography}{10}

\bibitem{kanda2019guided}
Naoyuki Kanda, Christoph Boeddeker, Jens Heitkaemper, Yusuke Fujita, Shota
  Horiguchi, Kenji Nagamatsu, and Reinhold Haeb-Umbach,
\newblock ``Guided source separation meets a strong asr backend:
  Hitachi/paderborn university joint investigation for dinner party asr,''
  2019.

\bibitem{CD2}
Gregory Sell and Daniel Garcia-Romero,
\newblock ``Speaker diarization with plda i-vector scoring and unsupervised
  calibration,''
\newblock in {\em 2014 IEEE Spoken Language Technology Workshop (SLT)}. IEEE,
  2014, pp. 413--417.

\bibitem{sell2015diarization}
Gregory Sell and Daniel Garcia-Romero,
\newblock ``Diarization resegmentation in the factor analysis subspace,''
\newblock in {\em 2015 IEEE International Conference on Acoustics, Speech and
  Signal Processing (ICASSP)}. IEEE, 2015, pp. 4794--4798.

\bibitem{fujita2019end}
Yusuke Fujita, Naoyuki Kanda, Shota Horiguchi, Kenji Nagamatsu, and Shinji
  Watanabe,
\newblock ``End-to-end neural speaker diarization with permutation-free
  objectives,''
\newblock {\em arXiv preprint arXiv:1909.05952}, 2019.

\bibitem{fujita2019end2}
Yusuke Fujita, Naoyuki Kanda, Shota Horiguchi, Yawen Xue, Kenji Nagamatsu, and
  Shinji Watanabe,
\newblock ``End-to-end neural speaker diarization with self-attention,''
\newblock in {\em 2019 IEEE Automatic Speech Recognition and Understanding
  Workshop (ASRU)}. IEEE, 2019, pp. 296--303.

\bibitem{tsvad_Medennikov}
Ivan Medennikov, Maxim Korenevsky, Tatiana Prisyach, Yuri Khokhlov, Mariya
  Korenevskaya, Ivan Sorokin, Tatiana Timofeeva, Anton Mitrofanov, Andrei
  Andrusenko, Ivan Podluzhny, and et~al.,
\newblock ``Target-speaker voice activity detection: A novel approach for
  multi-speaker diarization in a dinner party scenario,''
\newblock {\em Interspeech 2020}, Oct 2020.

\bibitem{horiguchi2020end}
Shota Horiguchi, Yusuke Fujita, Shinji Watanabe, Yawen Xue, and Kenji
  Nagamatsu,
\newblock ``End-to-end speaker diarization for an unknown number of speakers
  with encoder-decoder based attractors,''
\newblock {\em arXiv preprint arXiv:2005.09921}, 2020.

\bibitem{he21c_interspeech}
Maokui He, Desh Raj, Zili Huang, Jun Du, Zhuo Chen, and Shinji Watanabe,
\newblock ``{Target-Speaker Voice Activity Detection with Improved i-Vector
  Estimation for Unknown Number of Speaker},''
\newblock in {\em Proc. Interspeech 2021}, 2021, pp. 3555--3559.

\bibitem{yu2021m2met}
Fan Yu, Shiliang Zhang, Yihui Fu, Lei Xie, Siqi Zheng, Zhihao Du, Weilong
  Huang, Pengcheng Guo, Zhijie Yan, Bin Ma, et~al.,
\newblock ``M2met: The icassp 2022 multi-channel multi-party meeting
  transcription challenge,''
\newblock {\em arXiv preprint arXiv:2110.07393}, 2021.

\bibitem{AISHELL42021}
Yihui Fu, Luyao Cheng, Shubo Lv, Yukai Jv, Yuxiang Kong, Zhuo Chen, Yanxin Hu,
  Lei Xie, Jian Wu, Hui Bu, Xin Xu, and Jingdong~Chen Jun~Du,
\newblock ``Aishell-4: An open source dataset for speech enhancement,
  separation, recognition and speaker diarization in conference scenario,''
\newblock in {\em Interspeech}, 2021.

\bibitem{fan2020cn}
Yue Fan, JW~Kang, LT~Li, KC~Li, HL~Chen, ST~Cheng, PY~Zhang, ZY~Zhou, YQ~Cai,
  and Dong Wang,
\newblock ``Cn-celeb: a challenging chinese speaker recognition dataset,''
\newblock in {\em ICASSP 2020-2020 IEEE International Conference on Acoustics,
  Speech and Signal Processing (ICASSP)}. IEEE, 2020, pp. 7604--7608.

\bibitem{Raj2021ReformulatingDL}
Desh Raj and S.~Khudanpur,
\newblock ``Reformulating {DOVER-Lap} label mapping as a graph partitioning
  problem,''
\newblock {\em INTERSPEECH}, 2021.

\bibitem{landini2022bayesian}
Federico Landini, J{\'a}n Profant, Mireia Diez, and Luk{\'a}{\v{s}} Burget,
\newblock ``Bayesian hmm clustering of x-vector sequences (vbx) in speaker
  diarization: theory, implementation and analysis on standard tasks,''
\newblock {\em Computer Speech \& Language}, vol. 71, pp. 101254, 2022.

\bibitem{desplanques2020ecapa}
Brecht Desplanques, Jenthe Thienpondt, and Kris Demuynck,
\newblock ``Ecapa-tdnn: Emphasized channel attention, propagation and
  aggregation in tdnn based speaker verification,''
\newblock {\em arXiv preprint arXiv:2005.07143}, 2020.

\bibitem{park2019auto}
Tae~Jin Park, Kyu~J Han, Manoj Kumar, and Shrikanth Narayanan,
\newblock ``Auto-tuning spectral clustering for speaker diarization using
  normalized maximum eigengap,''
\newblock {\em IEEE Signal Processing Letters}, vol. 27, pp. 381--385, 2019.

\bibitem{snyder2015musan}
David Snyder, Guoguo Chen, and Daniel Povey,
\newblock ``Musan: A music, speech, and noise corpus,'' 2015.

\bibitem{aishell2}
J.~{Du}, X.~{Na}, X.~{Liu}, and H.~{Bu},
\newblock ``{AISHELL-2: Transforming Mandarin ASR Research Into Industrial
  Scale},''
\newblock {\em ArXiv}, Aug. 2018.

\bibitem{yoshioka2012generalization}
Takuya Yoshioka and Tomohiro Nakatani,
\newblock ``Generalization of multi-channel linear prediction methods for blind
  mimo impulse response shortening,''
\newblock {\em IEEE Transactions on Audio, Speech, and Language Processing},
  vol. 20, no. 10, pp. 2707--2720, 2012.

\bibitem{delcroix2014linear}
Marc Delcroix, Takuya Yoshioka, Atsunori Ogawa, Yotaro Kubo, Masakiyo Fujimoto,
  Nobutaka Ito, Keisuke Kinoshita, Miquel Espi, Takaaki Hori, Tomohiro
  Nakatani, et~al.,
\newblock ``Linear prediction-based dereverberation with advanced speech
  enhancement and recognition technologies for the reverb challenge,''
\newblock in {\em Reverb workshop}, 2014.

\bibitem{caroselli2017adaptive}
Joe Caroselli, Izhak Shafran, Arun Narayanan, and Richard Rose,
\newblock ``Adaptive multichannel dereverberation for automatic speech
  recognition.,''
\newblock in {\em Interspeech}, 2017, pp. 3877--3881.

\bibitem{drude2018nara}
Lukas Drude, Jahn Heymann, Christoph Boeddeker, and Reinhold Haeb-Umbach,
\newblock ``Nara-wpe: A python package for weighted prediction error
  dereverberation in numpy and tensorflow for online and offline processing,''
\newblock in {\em Speech Communication; 13th ITG-Symposium}. VDE, 2018, pp.
  1--5.

\bibitem{allen1979image}
Jont~B Allen and David~A Berkley,
\newblock ``Image method for efficiently simulating small-room acoustics,''
\newblock {\em The Journal of the Acoustical Society of America}, vol. 65, no.
  4, pp. 943--950, 1979.

\bibitem{zhang2017mixup}
Hongyi Zhang, Moustapha Cisse, Yann~N Dauphin, and David Lopez-Paz,
\newblock ``mixup: Beyond empirical risk minimization,''
\newblock {\em arXiv preprint arXiv:1710.09412}, 2017.

\bibitem{kingma2017adam}
Diederik~P. Kingma and Jimmy Ba,
\newblock ``Adam: A method for stochastic optimization,'' 2017.

\bibitem{fiscus2006rich}
Jonathan~G Fiscus, Jerome Ajot, Martial Michel, and John~S Garofolo,
\newblock ``The rich transcription 2006 spring meeting recognition
  evaluation,''
\newblock in {\em International Workshop on Machine Learning for Multimodal
  Interaction}. Springer, 2006, pp. 309--322.

\end{thebibliography}

\end{document}